\documentstyle[twoside,fleqn,espcrc2]{article}

\newcommand{\AmS}{{\protect\the\textfont2
  A\kern-.1667em\lower.5ex\hbox{M}\kern-.125emS}}
\newcommand{\be}{\begin{equation}}
\newcommand{\ee}{\end{equation}}
\newcommand{\bea}{\begin{eqnarray}}
\newcommand{\eea}{\end{eqnarray}}
\newcommand{\beff}{\beta_{\rm eff}}

\title{ Surface Tension, Surface Stiffness, and Surface Width
        of the 3-dimensional Ising Model on a Cubic Lattice
        \thanks{talk presented by K.\ Pinn} }
\author{M.\ Hasenbusch
        \address{CERN, Theory Division, CH-1211 Gen\`eve 23, Switzerland}%
        and
        K.\ Pinn
        \address{Institut f\"ur Theoretische Physik I,
        Universit\"at M\"unster, \\
        Wilhelm-Klemm-Str.\ 9, D-4400 M\"unster, Germany}}%
\begin{document}
\begin{abstract}
 We compute properties of the interface of the 3-dimensional Ising model
 for a wide range of temperatures and for interface extensions up to 64
 by 64. The interface tension $\sigma$ is obtained by integrating the
 surface energy density over the inverse temperature $\beta$.
 The surface stiffness coefficient $\kappa$ is determined. We also study
 universal quantities like $\xi^2 \sigma$ and $\xi^2 \kappa$.
 The behavior of the interfacial width on
 lattices up to $512 \times 512 \times 27$ is also investigated.
\end{abstract}
\maketitle
 Here we present a numerical study of 3-d Ising interfacial
 properties with focus on the precise determination of
 interface tension and interface stiffness over a wide range
 of temperatures.
 A detailed exposition of our work can be found
 in \cite{physica}.
\section{THE MODEL}\label{SECmodel}
 We consider a simple cubic lattice with extension
 $L$ in $x$- and $y$-direction and with extension $T=2D+1$
 in $z$-direction. The lattice sites $i=(i_x,i_y,i_z)$ have integer
 coordinates, and the $z$-coordinate runs from $-D$ to $+D$.
 The Ising model is defined by the partition function
\be
  Z = \sum_{ \{ \sigma_i = \pm 1\} } \exp(-\beta H) \, ,
\ee
 where
\be
  H= - \sum_{<i,j>} k_{ij} \sigma_i \sigma_j  \, .
\ee
 The lattice becomes a torus by imposing {\em geometrical}
 periodic boundary conditions in all three directions.
 For the Ising spins
 $\sigma_i$ we will use two different boundary conditions:
 Periodic boundary conditions are defined by letting
 $k_{ij}=1$ for all links $< \! i,j \!>$.
 To define antiperiodic boundary conditions in $z$-direction, we
 also set $k_{ij}=1$ with the exception of the
 links connecting the uppermost plane
 ($z=+D$) with the lowermost plane ($z=-D$). These links
 carry an antiferromagnetic factor $k_{ij}=-1$.
\section{INTERFACIAL PROPERTIES}\label{SECproper}
 We adopt the following definition of the {\em interfacial width}:
 A magnetization profile for lattice planes perpendicular to
 the $z$-direction is defined by $M(i_z) = L^{-2} \sum_{i_x,i_y} \sigma_i$.
 The antiperiodic boundary condition allows us to shift the configuration
 in $z$-direction such that the interface comes close to $i_z=0$.
 We introduce an auxiliary coordinate $z$ that assumes half-integer
 values (labelling positions between adjacent lattice layers
 perpendicular to the $z$-direction).
 Following \cite{mon90a}, a normalized magnetization
 gradient is defined as
\be
 \rho (z) =
 \frac{ M(z+ \frac12)-M(z-\frac12) }{ M(D)-M(-D) } \, .
\ee
 For a given spin configuration, the position of
 the interface is defined as $ \sum \rho (z) z $.
 The square of the interface width is then defined
 \cite{swrev,mon90a} as
\be
 W^2 = \bigl\langle \,
 \sum_z \rho(z) \, z^2 -  \bigl( \sum_z \rho(z) \, z \bigr)^2 \,
 \bigr\rangle \, .
\ee
 Especially on small lattices, fluctuations in the two bulk phases
 can deteriorate the results. Due to bubbles, $\rho(z)$ can
 be accidentially large even far away from the interface position.
 We therefore also consider the interface width
 measured on the configurations after a removal of all bubbles.

 We now turn to the definition of the {\em interface tension}.
 Let us assume that there is exactly
 one interface in the system with antiperiodic boundary conditions,
 and that there are no interfaces in the periodic system.
 Then the interfacial free energy is
 $ F_s = F_{a.p.} - F_{p.} + \ln T $,
 where $F_{a.p.}$ ($F_{p.}$) is the free energy ($= - \ln Z$)
 of the system with antiperiodic (periodic) boundary
 in $z$-direction. For a discussion of the case of several
 interfaces see \cite{physica}.
 The interface tension is defined as the limit
 $ \sigma= \lim_{L \rightarrow \infty} F_s / L^2 $.

 In general one has no direct access to the partition
 function in Monte Carlo simulations,
 except for not too large systems, cf.\ \cite{direct}.
 Note, however, that
\be
 \frac {\partial F_s}{\partial \beta} = \langle H \rangle_{a.p.}
                                      - \langle H \rangle_{p.}
 \equiv E_s \, .
\ee
 By integration one obtains
\be
 F_s(\beta) = F_s( \beta_0) + \int_{\beta_0}^{\beta}
 d \beta' \,  E_s(\beta') \, ,
\ee
 where $\beta_0 $ is arbitrary.
 Our approach is to compute by Monte Carlo simulation the surface energy
 for $\beta$-values ranging from the critical region around
 $\beta_c$ (= 0.221652(3) \cite{betac})
 up to $\beta_0 \equiv 0.6$.
 For large $\beta$ we can employ a low temperature expansion
 by Weeks et al.\ (published in
 an article by Shaw and Fisher \cite{ShawFisher}) to obtain
 $F_s(\beta_0)$.
 Note that it is also possible to start the integration
 at small $\beta$ \cite{physica}.
 (To the best of our
 knowledge, the integration method to obtain surface free energies
 was first used by B\"urkner and Stauffer \cite{kner83a}).

 In the theory of rough interfaces the
 {\em surface stiffness coefficient} $\kappa$ plays an important role.
 If by suitably chosen boundary
 conditions the low temperature interface is forced to make an
 angle $\theta$ with e.g.\ the $x$-axis, one defines
 (see e.g.\ \cite{priv}),
\be
 \kappa = \sigma(0) + \frac{d^2 \sigma}{d\theta^2} \vert_{\theta=0} \, .
\ee
 Capillary wave theory says that the long distance
 properties of the interface should by encoded in a
 2-dimensional Gaussian model with Hamiltonian
\be\label{GaussZ}
 H_0 = \frac{1}{2\beff} \sum_{<i,j>} (h_i - h_j)^2  \, ,
\ee
 and $\beff = 1/\kappa$.
 Long distance properties are most systematically studied via
 the block spin renormalization group.
 For the Gaussian model defined through eq.\ (\ref{GaussZ})
 one defines block spins $\Phi_I$ as averages over
 cubic blocks $I$ of size $L_B^2$, i.e.,
 $ \Phi_I = L_B^{-2} \sum_{i \in I} h_i $.
 We define the quantities
\be\label{aa1}
 A_{i,l}^{(0)} =
 \langle \frac 1 {l^2} \sum_{(I,J)} (\phi_I-\phi_J)^2 \rangle \, ,
\ee
 where $I$ and $J$ are nearest neighbors in the block lattice for
 $i=1$, and next-to-nearest neighbors for $i=2$.
 $l$ is the extension of the block lattice, i.e.\ $l = L/L_B$.
 For the Gaussian model, the $A$'s can be computed exactly.
 For the Ising model,
 block spin ``height variables'' $\bar h_I$ are defined as follows:
 The blocks $I$ are sets that are quadratic in
 $x-y$ direction with extension $L_B \times L_B$ and that extend
 through the {\em whole} lattice in $z$-direction.
 For every block, a magnetization profile
 and an interface position can be determined exactly as in
 the case of the full lattice. We define
 $ \bar h_I = $ interface position in block $I$.
 Note that the blocked height variables can also be defined
 ``with and without bubbles''.
 Blocked observables for the Ising interface are introduced
 analogously to eq.\ (\ref{aa1}), and are denoted
 by $A_{i,l}^{(\rm Ising)}$.
 For a rough Ising interface, we define an effective coupling
 $\beff$ as follows:
\be
 \beff = \lim_{L_B \rightarrow \infty}
 \frac{ A_{i,l}^{(\rm Ising)} }{ A_{i,l}^{(0)} } \, .
\ee
 Of course, we expect that the so defined $\beff$ does not depend
 on $i$ or $l$.
\section{MONTE CARLO RESULTS}\label{SECresults}

 We did simulations with antiperiodic boundary conditions
 in $z$-direction on lattices with $L=8,16,32,64$ for
 $\beta$-values ranging from the bulk critical region
 up to $\beta = 0.6$.
 For many $\beta$-values we made runs with different
 $D$ to control the effects of a finite thickness of the
 lattice. In total, we made more than
 250 different simulations with antiperiodic boundary
 conditions. Typically, we made 10000 measurements of
 several quantities, separated always by
 8 cluster updates with the Hasenbusch-Meyer
 cluster algorithm for Ising interfaces \cite{isiclust}.
 The simulations supplied us with
 a sufficiently dense grid of $\beta$-values for the energies
 $E_{a.p.}$.

 For most of the $\beta$-values, we fortunately
 did not have to do extra simulations to access the
 energies with periodic boundary $E_{p.}$. Instead we
 used the diagonal Pad\'e approximation of the low temperature
 series by Bhanot et al.\ \cite{epslow}.
 For smaller $\beta$-values we used the cluster Monte Carlo
 method to determine $E_{p.}$.

 In order to do the integration over $\beta$
 we interpolated the data with
 cubic splines. The integration was then started
 at $\beta=0.6$, where the integration constant can be
 safely taken from the low temperature series.

 Already for moderate surface extension $L$, the
 surface free energy was found to behave with very good precision like
 $ F_s = C_s + \sigma' \, L^2 $.
 It was therefore natural to identify the coefficient $\sigma'$
 with the surface tension $\sigma$.

 The results for the free energies for $L=8$, $16$, $32$ and $64$
 were then used to make fits with $ F_s = C_s + \sigma \, L^2 $
 in order to obtain estimates for the surface tension $\sigma$.
 In table \ref{tab1} we display a few of our results.
 Our results show a significant deviation from a
 prediction by Shaw and Fisher \cite{ShawFisher} based
 on an analysis of the low temperature series.
 A detailed comparison will be published elsewhere \cite{fishcomp}.
\begin{table*}[hbt]
\setlength{\tabcolsep}{1.25pc}
\newlength{\digitwidth} \settowidth{\digitwidth}{\rm 0}
\catcode`?=\active \def?{\kern\digitwidth}
\caption{A few results for $\sigma$ and for $\beff$ }
\label{tab1}
\begin{tabular}{|c||c|c|c|c|c|}
\hline
$\beta$  & 0.2275    & 0.2327    & 0.2391    & 0.3000     & 0.40236   \\
\hline
$\sigma$ & 0.0146(1) & 0.0319(1) & 0.0555(1) & 0.30284(8) & 0.67988(6) \\
\hline
\hline
$\beta$  & 0.240     & 0.275      & 0.330      & 0.35    & 0.37     \\
\hline
$\beff$  & 16.7(5)   & 4.65(4)    & 1.93(2)    & 1.52(2) & 1.20(2)  \\
\hline
\end{tabular}
\end{table*}

 We fitted our results for $\sigma$
 to the critical law $\sigma = \sigma_0 t^{\mu}$, using both
 of the two definitions $t=1-\beta_c/\beta$ and
 $t=\beta/\beta_c - 1$. We also varied the interval over
 which the $\beta$-dependence of $\sigma$ was fitted.
 The fits were always done using four different $\beta$-values.
 The results based on
 the two different definitions of $t$ were statistically
 incompatible, showing that one is still not close enough
 to criticality. However, taking systematic effects into account,
 we consider our results consistent with $\mu \approx 1.26$.
 The results for the critical amplitude
 $\sigma_0$ show even stronger dependency on the type of the fit,
 and we can not say very much more than that $\ln \sigma_0$ is probably
 something between $0.2$ and $0.4$.

 In order to study the behavior of
 the product $\xi^2 \sigma$ we tried to
 determine the correlation length $\xi$ from the simulations of the system
 with periodic boundary conditions. We defined $\xi$
 via the decay of the connected 2-point correlation function of
 the absolute value of the time-slice magnetization.
 Our results are nicely consistent with low temperature series
 \cite{Arisue}.

 We measured the block spin correlation functions
 $ A_{i,l}^{(\rm Ising)} $ and studied the quantities
 $ \beta_{\rm eff}^{i,l}(\lambda) =
 A_{i,l}^{(\rm Ising)}(L_B=\lambda) / A_{i,l}^{(0)}(L_B=\infty) $
 (measured on the configurations with the bubbles removed).
 The values for the different $i,l$ and $\lambda$'s
 turned out to be fairly
 consistent within the statistical accuracy.
 Our estimates for $\beff$ were then determined
 by averaging over $\beta_{\rm eff}^{i,l}$ with
 $i=1$ and $i=2$. Some of our results are shown in
 table \ref{tab1}.

 In figure \ref{fig1}, we show our results for two combined quantities,
 namely $\xi^2 \sigma$ and $\xi^2 \kappa$.
 In the product $\xi^2 \sigma$, the exponents $\mu$ and $-2\nu$
 of the reduced temperature $t$ should cancel, and we expect
 that this product should be fairly constant in a neighborhood of the
 critical point. The full line in the figure was obtained by combining
 our $\sigma's$ from the integration method with the correlation lengths
 as obtained from the Pad\'e.

 Since we do not know the error of the Pad\'e approximation of the
 low temperature series we base our error estimate
 for this quantity
 on our error bars for the measured correlation length
 and on the statistical errors on the surface tension $\sigma$.
 We estimate the relative precision of our results
 for $\xi^2 \sigma$ to be around 5 per cent for the smaller
 $\beta$'s, certainly better in the large $\beta$ region.
 This takes into account statistical errors only. There might
 also be systematic errors (due to too small $L$'s)
 in the surface tension close to
 the critical point. They might be responsible for another
 5 percent relative uncertainty.
 The points with error bars in the figure show the product
 $\xi^2 \kappa$. The plot shows that in the critical limit
 the surface stiffness becomes the same as the interface tension.
 This is a consequence of the restoration of rotational symmetry
 at the bulk critical point.
\begin{figure}[htb]
\vspace{60mm}
\caption{Results for $\xi^2 \sigma$ and $\xi^2 \kappa$}
\label{fig1}
\end{figure}
 Using our results for both $\xi^2 \sigma$
 and $\xi^2 \kappa$, we estimate that in the limit $\beta \rightarrow \beta_c$
 both quantities have the limiting value $R_- = 0.90(5)$.

 In the theory of critical wetting, the quantity
 $ \omega(\beta) = 1 / (4 \pi \kappa \xi^2) $
 plays an important role.
 In a paper of Fisher and Wen \cite{FisherWen} this quantity is
 estimated over a wide range of temperatures.
 A comparison of our data with their theoretical prediction
 will be published elsewhere \cite{fishcomp}.

 When $\beta$ approaches the roughening coupling
 $\beta_R$, Kosterlitz-Thouless theory
 states that $\beff \rightarrow 2/\pi$. Using the estimates
 $\beta_R = 0.4074(3)$ \cite{martinthesis}, and
 $\xi(\beff) = 0.3163$ (from the Pad\'e that here certainly is reliable),
 we find a ``KT value'' of $\xi^2 \kappa$ which is 0.1572.

 In order to demonstrate the efficiency of the Hasenbusch-Meyer
 algorithm we redid the surface width computation of Mon et al.\
 \cite{mon90a} at $ \beta = \beta_c/0.8 = 0.2771 $ on lattices of
 size $L \times L \times 27$, with $L=32,64,\dots,512$.
 We performed fits of the data using the ansatz
\be
 W^2 = \mbox{const} + \frac{\beff}{2\pi} \ln L
\ee
 that is motivated by Kosterlitz-Thouless theory of a rough interface.
 When the $L=32$ data were excluded, the fits where quite convincing.
 From our analysis we arrive at an estimate $\beff=4.3(2)$.
 This result is nicely consistent with the $\beff$ as obtained from
 the renormalization group quantities $A$.

\smallskip\noindent
 M.H.\ would like to thank the Deutsche Forschungsgemeinschaft for
 financial support.


\begin{thebibliography}{9}
%
\bibitem{physica}
 M.\ Hasenbusch and K.\ Pinn,
 \newblock preprint M\"unster MS-TPI-92-24, to be published in Physica A.
%
\bibitem{mon90a}
 K.K.\ Mon, D.P.\ Landau and D.\ Stauffer,
 \newblock Phys.\ Rev.\ B42, 545 (1990).
%
\bibitem{swrev}
 R.H.\ Swendsen,
 \newblock Phys.\ Rev.\ B15, 542 (1977).
%
\bibitem{direct}
 M.\ Hasenbusch,
 \newblock preprint Kaiserslautern KL-TH 16/92, to be published
  in J.\ Physique I.
%
\bibitem{betac}
 C.F.\ Baillie, R.\ Gupta, K.A.\ Hawick and G.S.\ Pawley,
 \newblock Phys.\ Rev.\ B45, 10438 (1992).
%
\bibitem{ShawFisher}
 L.J.\ Shaw and M.E.\ Fisher,
 \newblock Phys.\ Rev.\ A39, 2189 (1989).
%
\bibitem{kner83a}
 E.\ B\"urkner and D.\ Stauffer,
 \newblock Z.\ Phys.\ B53, 241 (1983).
%
\bibitem{priv}
 V.\ Privman,
 \newblock Phys.\ Rev.\ Lett.\ 61, 183 (1988).
%
\bibitem{isiclust}
 M.\ Hasenbusch and S.\ Meyer,
 \newblock Phys.\ Rev.\ Lett.\ 66, 530 (1991).
%
\bibitem{epslow}
 G.\ Bhanot, M.\ Creutz and J.\ Lacki,
 preprint IASSNS-HEP-92/42.
%
\bibitem{Arisue}
H.\ Arisue and T.\ Fujiwara,
 \newblock Nucl.\ Phys.\ B285 [FS 19], 253 (1987).
%
\bibitem{FisherWen}
 M.E.\ Fisher and H.\ Wen,
 \newblock Phys.\ Rev.\ Lett.\ 68, 3654 (1992).
%
\bibitem{fishcomp}
 M.\ Hasenbusch and K.\ Pinn, in preparation.
%
\bibitem{martinthesis}
 M.\ Hasenbusch,
 \newblock PhD-thesis, Universit\"at Kaiserslautern, 1992.
\end{thebibliography}
\end{document}